\documentclass[prl,twocolumn,showpacs]{revtex4}
\usepackage[dvips]{graphicx}
\usepackage{latexsym}
\usepackage{bm}
\begin{document}
\newcommand{\fig}[2]{\includegraphics[width=#1]{#2}}
\newcommand{\pprl}{Phys. Rev. Lett. \ }
\newcommand{\pprb}{Phys. Rev. {B}}
\newcommand{\be}{\begin{equation}}
\newcommand{\ee}{\end{equation}}
\newcommand{\bea}{\begin{eqnarray}}
\newcommand{\eea}{\end{eqnarray}}
\newcommand{\nn}{\nonumber}
\newcommand{\la}{\langle}
\newcommand{\ra}{\rangle}
\newcommand{\dg}{\dagger}
\newcommand{\upa}{\uparrow}
\newcommand{\dna}{\downarrow}
\newcommand{\bisco}{Bi$_2$Sr$_2$CaCu$_2$O$_{8+\delta}$}
\newcommand{\oxychloride}{Ca$_{2-x}$Na$_x$CuO$_2$Cl$_2$}
\newcommand{\bione}{Bi$_{2-x}$Pb$_x$Sr$_2$CuO$_{y}$}
\newcommand{\tj}{$t$-$J$ }
\title{Correlating off-stoichiometric doping with nanoscale electronic
disorder and quasiparticle interference pattern in high-$T_c$
superconductor \bisco}

\author{Sen Zhou, Hong Ding, and Ziqiang Wang}
\affiliation{Department of Physics, Boston College, Chestnut Hill, MA 02467}

\date{\today}

\begin{abstract}

A microscopic theory is presented for the observed
electronic disorder in superconducting \bisco.
The essential phenomenology is shown to be
consistent with the existence of two types of interstitial oxygen dopants:
those serving primarily as charge reserviors and those close to the apical
plane contributing both carriers and electrostatic potential
to the CuO$_2$ plane. The nonlinear screening of the latter
produces nanoscale variations in the doped hole
concentration, leading to electronic inhomogeneity.
Based on an unrestricted Gutzwiller approximation of the
extended \tj model, we provide a consistent
explanation of the correlation between the observed dopant
location and the pairing gap and its spatial evolutions.
We show that the oxygen dopants are the primary cause of
both the pairing gap disorder and the quasiparticle interference pattern.

\typeout{polish abstract}
\end{abstract}


\maketitle

Remarkable electronic inhomogeneities have been observed by
scanning tunneling microscopy (STM) in high-$T_c$ superconductors
\bisco \cite{pan,howald,madhavan,McElroy,fang},
\oxychloride \cite{kohsaka,hanaguri}, and \bione \cite{mashima}
over a wide range of doping.
The hallmark of the inhomogeneity is the disordered,
nanometer scale variation of the pairing
energy gap and its anti-correlation with that of the coherence
peak height in the local density of states (LDOS). The origin
of the electronic disorder has been the focus of several theoretical
studies \cite{wangetal,dhlee,martin,su,nunner}.
It was emphasized \cite{pan,wangetal,dhlee} that
off-stoichiometric doping a Mott insulator, such as the
high-T$_c$ cuprates, creates inherently inhomogeneous
electronic states associated with the interstitial or substitutional
dopant atoms. Ionized, these off-plane dopants act as disordered centers of
nonlinearly screened electrostatic potential, leading to a spatially
inhomogeneous distribution of the local doping concentration (LDC) and hence
that of the local pairing gap. In a short coherence length superconductor
described by the short-range resonance valence bond theory
\cite{anderson,kr}, the local pairing gap is anti-correlated with the
LDC \cite{wangetal,dhlee}.

The notion of dopant induced electronic disorder is supported and further
advanced by the recent STM experiments of McElroy et.al. \cite{McElroy}.
Identifying the spectral peak around $-0.96$eV in the LDOS
with the presence of a local excess oxygen atom, the correlation
between the dopant location
and the gap inhomogeneity has been established ubiquitously.
However, the observed dopants reside close to
the regionos of large pairing gap.
This is counter-intuitive since the negatively charged
oxygen ions are expected to attract nearby holes and
create a higher LDC with a smaller pairing gap
\cite{wangetal,dhlee}. This discrepancy promotes the idea
that dopant induced local structural distortions may play a more important
role than potential disorder \cite{McElroy,uchida} and
phenomenological theories in which the dopants serve as
large pairing centers \cite{nunner}. The origin of the dopant induced
electronic disorder has remained a central unresolved issue.

In this paper, we show that, while dopant induced structural effects are
present, the main cause for the electronic disorder
is the dopant potential induced LDC modulations. A clue as to how the
latter can be consistent with the dopant locations comes from
the following observations made on the experiments of McElroy et.al.
\cite{McElroy}: (1) The observed dopants associated
with the $-0.96$eV peak cannot account for the total
number of oxygen dopants. For example,
in the underdoped sample with $\bar\Delta=65$meV,
the observed dopant density is $2.7\%$.
Even if every oxygen dopant is fully ionized,
the average doping would only be $5.5\%$, which
is much smaller than the expected value around $11\%$ \cite{McElroyPRL05}.
Thus, a substantial number of the dopants has not been located,
too large to be accounted for by
the uncertainty in the doping process.
(2) The correlation between the observed dopants and the pairing gap
is weak and well defined only in the statistical sense.
An appreciable number of them can be found in the
smaller gap regions or to straddle
the boundaries between the small and large gap regions.
Thus, it is unlikely that the observed dopants strongly and directly affect
the {\it local} electronic structure.
(3) The in-gap, low energy states in the
LDOS have their weight concentrated in regions away from the identified
dopants. This suggests that these electronic
states which are encoded with the quasiparticle interference modulations
\cite{McElroy} are likely localized or pinned by additional confining
potentials in regions away from the identified dopants.
 
Based on these observations, we conjecture that there are two types (A and B)
of interstitial oxygen dopants in \bisco. The type-B dopants serve primarily
as charge reserviors. They only couple weakly to the CuO$_2$ plane.
These near nonbonding oxygen orbitals give rise to the $-0.96$eV peak in
the LDOS \cite{McElroy}. Recent ARPES measurements show that the dopant induced
states near $-0.96$eV have $B_1$ symmetry and do not mix with
the doped holes residing in the planar orbitals \cite{hongdopant}. This
further supports the identification of the former with the type B-dopants.
The type-A dopants, on the other hand, strongly affect the local electronic
structure in the CuO$_2$ plane. Their electrostatic potential enhances the LDC
which in turn pushes the orbital energy of type-A dopants
into the broad valence band spectra below $-1.2$eV not yet
accessible by STM probes. One possibility is that
the type-B dopants sit above the BiO plane while
type-A dopants come close to the apical SrO layer. We will discuss
the microscopic origin of the latter at the end.

To support this physical picture,
we extend the \tj model to include the oxygen dopant potential,
\begin{eqnarray}
H=
&-&\sum_{i\neq j}t_{ij}Pc_{i\sigma}^\dagger c_{j\sigma}P
+J\sum_{\langle i,j\rangle}({\bf S_i}\cdot{\bf S_j}
-{1\over4}\hat n_i\hat n_j)
\nonumber \\
&+&\sum_{i\neq j}\hat n_i V_{ij}^c \hat n_j
-\sum_{i} (V_{i}^A+V_{i}^B) (1-\hat n_{i}).
\label{h}
\end{eqnarray}
Here $c_{i\sigma}^\dagger$ creates an electron
that hops between near neighbors of the Cu square lattice
via $t_{ij}$. Repeated spin indices are summed and
$\hat n_i=c_{i\sigma}^\dagger c_{i\sigma}$ is the density operator.
The LDC is given by
$x_i=1-n_i$, $n_i=\langle c_{i\sigma}^\dagger c_{i\sigma}
\rangle$. The average doping will be denoted
as $\delta=(1/N_s)\sum_i x_i$ on a lattice of $N_s$ sites.
The second line in Eq.~(\ref{h}) describes
the long-range Coulomb interactions
between the electrons in the plane and
between the in-plane doped holes and the two-types of off-plane dopants,
$V_{ij}^c=V_c/\vert r_i-r_j\vert$ and
\begin{equation}
V_{i}^{A(B)}=\sum_{\ell_{A(B)}=1}^{N_{A(B)}}{2 V_{A(B)}\over
\sqrt{\vert r_i-r_{\ell_{A(B)}})\vert^2+d_{A(B)}^2}}
\label{v}
\end{equation}
where $d_A$, $d_B$ are the setback distances
and $N_A$, $N_B$ the number of type-A and type-B dopants
respectively, $\delta=2(N_A+N_B)/N_s$.

Eq.~(\ref{h}) describes doped Mott insulators because of the
projection operator $P$ that removes double occupation.
The projection is most conveniently implemented using the Gutzwiller
approximation by the statistical
weighting factors multiplying the coherent states, thus renormalizing
the hopping and the exchange parameters \cite{zhang}. Since the dopants break
translation symmetry, we extend the approach to the
disordered case by the renormalization $t_{ij}\to g_{ij}^t t_{ij}$
and $J\to g_{ij}^J J$ where the Gutzwiller factors
\begin{equation}
g_{ij}^t=\sqrt{4x_i x_j\over (1+x_i)(1+x_j)},\quad
g_{ij}^J={4\over (1+x_i) (1+x_j)}
\label{gfactor}
\end{equation}
depend on the local doping at the sites
connected by the hopping and the exchange processes.
The exchange term is decoupled in terms of the
bond $\chi_{ij}=\langle c_{i\sigma}^\dagger c_{j\sigma}
\rangle$ and the pairing $\Delta_{ij}=\langle \epsilon_{\sigma\sigma^\prime}
c_{i\sigma}c_{j\sigma^\prime}\rangle$ fields, leading to
a renormalized mean-field Hamiltonian,
\begin{eqnarray}
H_{\rm GA}=
&-&\sum_{i\neq j}g_{ij}^t t_{ij}c_{i\sigma}^\dagger c_{j\sigma}
+\sum_i\varepsilon_{i}c_{i\sigma}^\dagger c_{i\sigma}
\nonumber \\
&-&{1\over4}J\sum_{\langle i,j\rangle}g_{ij}^J
[\Delta_{ij}^*\epsilon_{\sigma\sigma^\prime} c_{i\sigma} c_{j\sigma^\prime}
+\chi_{ij}^*c_{i\sigma}^\dagger c_{j\sigma}+{\rm h.c.}]
\nonumber \\
&+&{1\over4}J\sum_{\langle i,j\rangle}
g_{ij}^J(\vert\Delta_{ij}\vert^2+\vert\chi_{ij}\vert^2)
-\sum_i\lambda_i n_i.
\label{hga}
\end{eqnarray}
The local energy for the electrons
is given by $\varepsilon_{i}=V_{\rm sc}(i)+\lambda_i-\mu_f$, where
$\mu_f$ is the chemical potential and $\lambda_i$ is a fugacity
that, together with the last term, ensures the
equilibrium condition under local occupation
$\langle c_{i\sigma}^\dagger c_{i\sigma}\rangle=n_i$ \cite{chlietal}.
$V_{\rm sc}(i)$ is the screened Coulomb potential,
\begin{equation}
V_{sc}(i)= V_i^A+V_i^B+V_c\sum_{j\neq i}{x_i-\delta\over\vert r_i-r_j\vert}.
\label{vsc}
\end{equation}
This is the driving force of the electronic disorder through local
doping variations. The correlation of the latter to the local
pairing gap disorder is caused by the Gutzwiller factor $g^t$
in Eq.~(\ref{gfactor}) that modulates the kinetic energy locally.
Minimizing the ground state energy of Eq.~(\ref{hga}), we obtain the
self-consistent equations for the set of parameters
$(\Delta_{ij},\chi_{ij},\lambda_i,x_i,\mu_f)$, which are solved iteratively
for a given average doping $\delta$.

We present our results for systems of $32\times32$ sites.
We use $t_{ij}=(0.48,-0.16,0.05,0.05,-0.05)$eV and $J=0.08$eV
such that the quasiparticle dispersion in Eq.~(\ref{hga}) agrees
with that measured by angle-resolved photoemission \cite{arpesband}.
For simplicity, half of the dopants are taken to be B-type and
the other half A-type distributed randomly with $d_A=1a$. The bare
electrostatic potentials are set to $V_A=V_c=0.5$eV, and $V_B=0$ for the weak
coupling between B-dopants and doped holes.
Note that the locations of type-A and type-B dopants are
{\it naturally anticorrelated }over the average interdopant distance
$a\sqrt{2/\delta}$. The spatial distribution
of the LDC $x_i$ and the d-wave pairing order parameter $\Delta_d(i)$
are shown as 2D maps in Figs.~1a and 1b for a typical system at
$\delta=10.2\%$. The projected dopant locations are superimposed.
The A-dopants are ubiquitously correlated
with the LDC. Due to the local pairing nature of the short-range RVB state,
$x_i$ is in turn strongly anti-correlated with $\Delta_d(i)$
\cite{wangetal}. Notice that the modulations
induced by the A-dopants leaves the B-dopants unwittingly correlated
overall with the low doping and strong pairing regions.
The LDOS is calculated from the projected retarded Green's function
in the Gutzwiller approximation \cite{chlietal},
${\rm LDOS}(i,\omega)=-2{\rm Im}g_{ii}^tG(i,i,\omega+i0^+)$.
To reduce finite size effects, we average over different boundary
conditions corresponding to $20\times20$ supercells.
As in the STM experiments, the local tunneling gap $\Delta_T$ is extracted
from the coherence peak position at positive bias
in the LDOS. In Fig.~1c, the tunneling gap map is shown with the dopant
locations. Evidently, $\Delta_T$ is small near the
A-dopants where the LDC is high. Notice however, that
the B-dopants are found with high statistics in and around the regions
where the tunneling gap is large, consistent with their identification
with those observed by STM \cite{McElroy}.
To elucidate the correlation between the
dopants and electronic disorder, we calculate the normalized cross-correlation
function between the dopants and the tunneling gap,
$$
C_{\Delta_T-O_{A(B)}}(r_j)={\sum_i\delta \Delta_T(r_i) \delta O_{A(B)}
(r_i+r_j)\over\sqrt{\sum_i[\delta\Delta_T(r_i)]^2\sum_j[\delta O_{A(B)}
(r_j)]^2}}
$$
where $\delta f(r_i)=f(r_i)-\langle f\rangle$. The dopant locations are modeled
by Lorentzians of width $2a$ \cite{McElroy}.
Fig.~1d shows that, while $\Delta_T$ is strongly
anticorrelated with A-dopant locations ($C_{\Delta_T-O_A}(0)\sim -0.6$),
the B-dopants (observed by STM) are positively correlated with the local
tunneling gap, and moreover, the correlation is significantly weaker,
$C_{\Delta_T-O_A}(0)\sim 0.3$, in excellent agreement with experiments
\cite{McElroy}.
\begin{figure}
\begin{center}
\fig{3.3in}{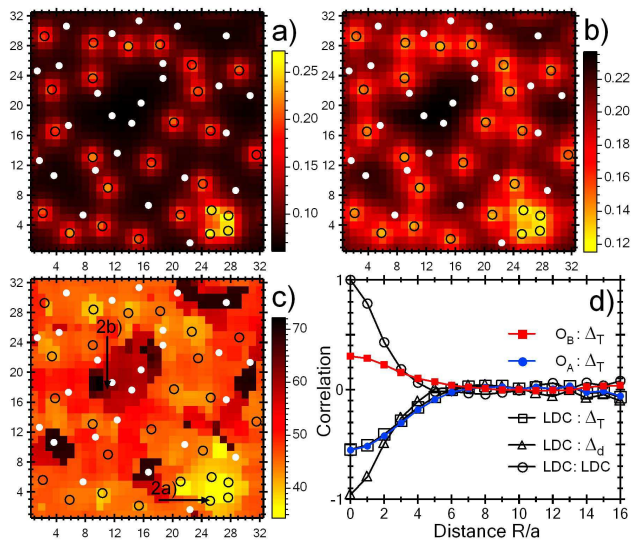}
\vskip-2.4mm
\caption{
Doping and pairing disorder on a $32\times32$ system at $\delta=10.2\%$.
2D maps are shown for the LDC $x_i$ (a), dimensionless d-wave pairing order
parameter $\Delta_d$ (b), and tunneling gap $\Delta_T$ in meV
(c) with the projected type-A (open black circle) and type-B (white solid
circle) dopant locations superimposed.
(d) Correlation functions among the dopant location, $\Delta_T$,
$x_i$, and $\Delta_d$.}
\label{fig1}
\vskip-9mm
\end{center}
\end{figure}

The large LDC variation is not inconsistent with
the STM integrated LDOS variation that diminishes only when
integrated up to $-0.6$eV to $-0.9$eV, where large incoherent background
and the spectral weight associated with the
dopants contribute appreciably \cite{McElroy}.
It is a subtle task to relate quantitatively the integrated LDOS
or the topography to the LDC since the former are obtained in
the constant tunneling current mode where the tip to sample distance
changes significantly, leading to reduced spatial variations of the
spectral weight \cite{pan,wangetal}.
\begin{figure}
\begin{center}
\fig{2.8in}{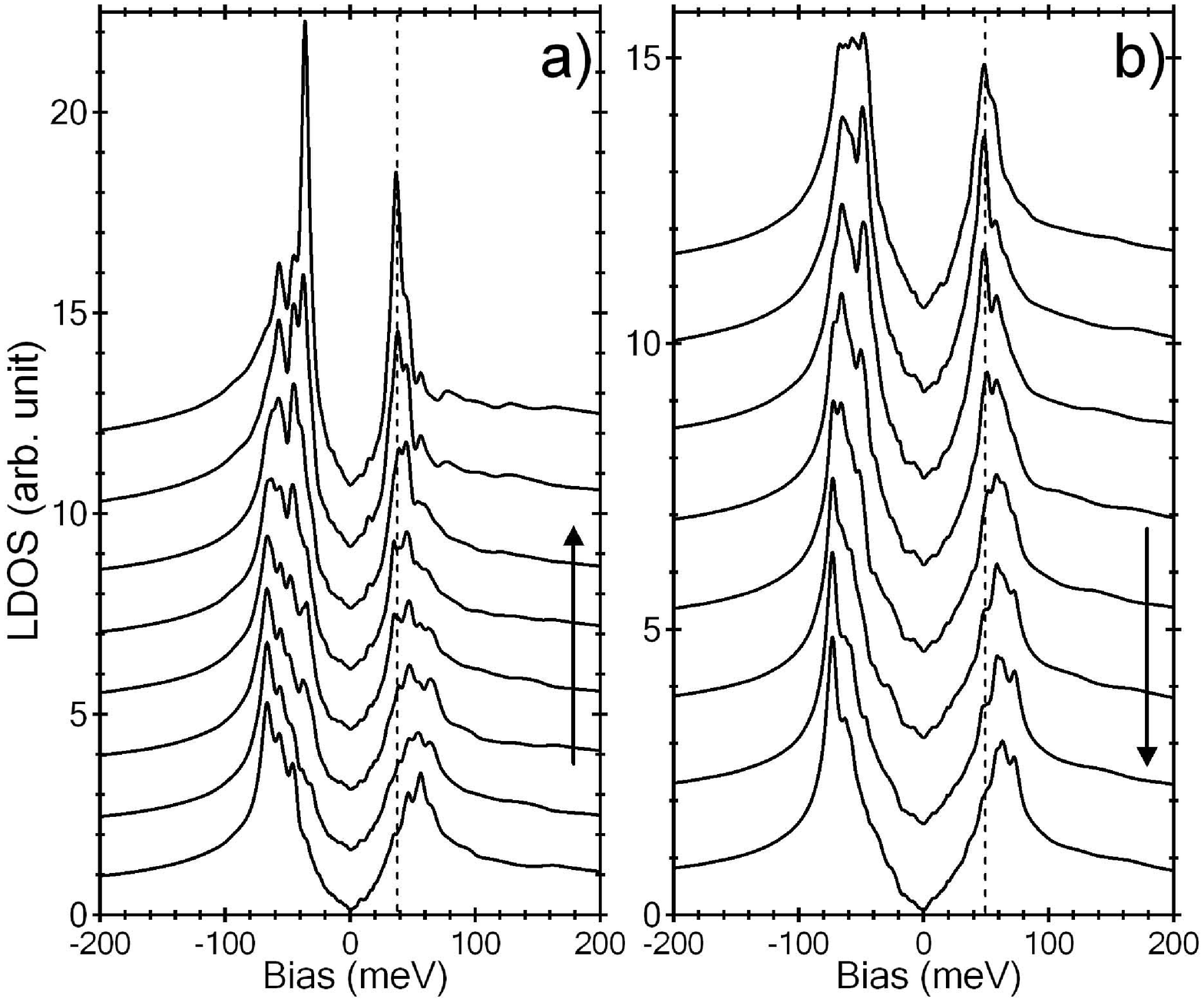}
\vskip-2.4mm
\caption{Evolutions of the LDOS along two line cuts
(shown in Fig.~1c) from small to average (a) and average to large (b)
gap regions. Dashedlines are guide to eye.
}
\label{fig2}
\vskip-9mm
\end{center}
\end{figure}

This form of dopant induced electronic disorder can describe
the basic properties of the inhomogeneous low energy states
observed experimentally. In Fig.~2, we present the LDOS along the two
line-cuts indicated in Fig.~1c.
Clearly, larger (smaller) gap regions are associated with smaller (larger)
coherence peaks in agreement with
STM \cite{pan,howald,madhavan,McElroy,fang}.
Note that the line-cut passes through
locally highly overdoped regions with $x_i>25\%$.
In a uniform system, the pairing gap
would be vanishingly small and the system essentially in the normal state
at such high doping levels.
Fig.~2a and Fig.~1c show, however, that the smaller local gap is still sizable.
That overdoped regions of sizes not exceeding the coherence
length have gap values considerably above those in a clean sample at the
corresponding dopings was first pointed out in Ref.\cite{wangetal},
and is likely a manifestation of the proximity effect due to the surrounding
larger gap regions. Indeed, Fig.~2 shows that as we move away from the
small (large) gap region, a larger (smaller) gap emerges and evolves
progressively stronger until it becomes the dominant gap as the line cut enters
the lower (higher) doping region.
This feature has been observed recently by STM with
high energy resolutions \cite{fang}.

Next we show the A-dopants, the culprit of strong gap disorder, also
serve as the primary cause of the low-energy quasiparticle interference
modulations \cite{McElroy03}. In Fig.~3a, the 2D maps of the
${\rm LDOS}(i,\omega)$ is plotted at fixed energies $\omega$,
showing different interference patterns.
The maxima of the spatial modulations are preferentially centered
around the A-dopant sites, thus away from the B-dopants, consistent
with STM observations. This turns out to be true for all low
energy states.
In Fig.~3c, we plot the {\it local} correlation between the LDOS
and the dopant locations as a function of bias voltage.
For both positive and negative energies, the correlations are positive and
strong (reaching $0.9$) with A-dopants and negative (anticorrelated)
and weak (reaching $-0.3$) with B-dopants.
The peak and dip around $\pm70$meV are related to the
Van Hove-like feature in the LDOS spectra seen in Fig.~2.
Just as in the case of dopant-gap correlations, the strong
{\it correlation}
of the interference modulations with A-dopants
produces a weak {\it anticorrelation} with the B-dopants as
in STM experiments \cite{McElroy}.
To further illustrate the interference pattern,
the Fourier transform of the quasiparticle LDOS
are shown in Fig.~3b at the corresponding energies.
The dominant interference wavevector ${\bf q}_1$ (marked by arrows)
connecting the tips of the Fermi arcs \cite{qhwang} are clearly seen
to disperse with energy near $(0,\pm 2\pi/4a)$ and $(\pm 2\pi/4a,0)$,
while the ${\bf q}_7$ connecting two tips of the same arc shows the
opposite dispersion with energy, in remarkable agreement
with STM observations \cite{McElroyPRL05,McElroy03}.
\begin{figure}
\begin{center}
\fig{3.3in}{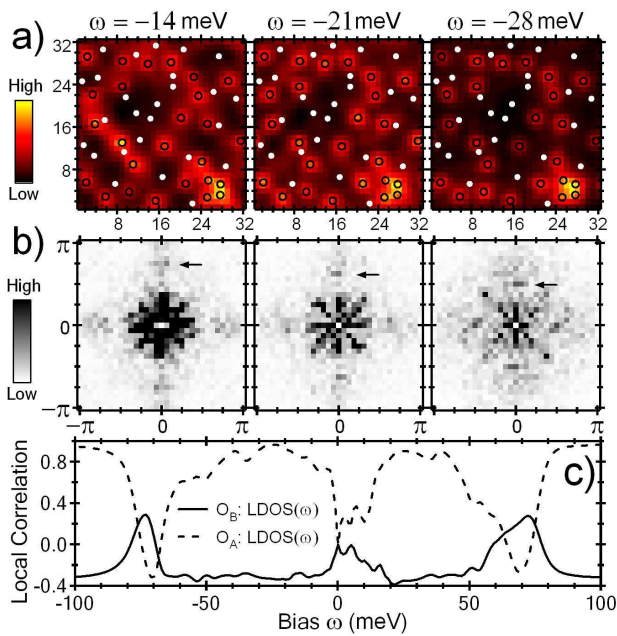}
\vskip-2.4mm
\caption{
Electron LDOS maps (a) and the Fourier transform of the quasiparticle
interference pattern (b) at low energies.
(c) Local correlations of ${\rm LDOS}(i)$ and the dopant
locations as a function of bias energy.
}
\label{fig3}
\vskip-9mm
\end{center}
\end{figure}

We now provide a possible microscopic origin of
the type-A dopants. It is well known that Bi-based cuprates have
a natural tendency toward Bi:Sr nonstoichiometry, i.e.
a fraction of the Bi comes to the SrO apical plane and
replaces the Sr in order to form the crystal structure
\cite{watanabe}. This creates the so-called A-site disorder
\cite{eisaki,uchida}. Since the trivalent Bi$^{3+}$ replacing Sr$^{2+}$
creates an excess positive charge locally, it naturally attracts the
negatively charged interstitial oxygen dopants to the vicinity of the
apical plane, forming the A-dopants. The doping levels derived
from the observed type-B dopants alone are about $5\sim6\%$ less than the
expected values in the samples studied by McElroy, et.al. \cite{McElroy},
suggesting about $5\sim6\%$ of the doped holes must come from the
$\sim3\%$ ``missing'' A-dopants. This is reasonably
consistent with the typical Bi:Sr nonstoichiometry in Bi2212 where about
$5\%$ of the Sr in each apical plane is replaced by Bi \cite{watanabe}.
Recently, the effect of A-site
disorder on $T_c$ has been studied systematically by controlled trivalent
(Ln$^{3+}$) substitution of Sr$^{2+}$ in
Bi$_2$Sr$_{2-y}$Ln$_y$CuO$_{6+\delta}$ where
Ln=La, Pr, Nd, Sm, Eu, Gd, and Bi \cite{eisaki,uchida}.
We propose that the STM experiments
be carried out on Bi2212 samples with controlled trivalent substitution
of Sr in the apical plane. Our theory predicts that the density of the
observable type-B dopants would decrease with increasing Ln$^{3+}$
substitution, while that of the type-A dopants increases, leading to
stronger electronic disorder.

We have shown that the electrostatic potential
of the off-stoichiometric A-dopants can be the primary cause of the electronic
disorder and quasiparticle interference modulations observed
in \bisco. The electronic inhomogeneity
in our theory is driven by that of the kinetic energy or the coherence of
doped holes in a doped Mott insulator. Incoherent excitations
beyond the Gutzwiller approximation and the dopant induced structural
distortions can also contribute to the electronic disorder.
We expect such dopant induced electronic disorder in all doped cuprate
superconductors, with specific properties dependent on the
dopant locations and the crystal field environment,
interstitial or substitutional, ordered or disordered.
The off-plane dopant structure together with the role
of the apical oxygen may account for the varying properties of the
cuprates that share, otherwise, identical CuO$_2$ planes.

We thank J.~C. Davis, P. Hirschfeld, W. Ku, D.-H. Lee, P.~A. Lee, C. Li,
S.~H. Pan, and S. Uchida for discussions. This work is supported
by DOE grant DE-FG02-99ER45747, ACS 39498-AC5M,
and NSF DMR-0353108. ZW thanks the KITP at UCSB for hospitality
and acknowledges the support of NSF grant PHY94-07194.

\end{document}